\documentstyle{article}
\newcommand{\lettersize}{\large \baselineskip=0.8cm}
\newcommand{\fig}[4]
{
     \LARGE
     \noindent
     \unitlength=1mm
     \begin{picture}(140,130)
     \put(0,0){
       \psfig{figure=#2,width=#3,height=#4}
     }
     \end{picture}
     \vfill
     \normalsize
     {\tt
       \noindent
       Figure #1 \\
       \figauth \\
     }
   \noindent
}
\newcommand{\figpage}[2]{\fig{#1}{#2}{140mm}{120mm}}
\newcommand{\xfigpage}[2]{\fig{#1}{#2}{120mm}{120mm}}
\newcommand{\confpage}[2]{\fig{#1}{#2}{140mm}{70mm}}
\newcommand{\figauth}{F. Schmid, C. Stadler, H. Lange}
\begin{document}
\lettersize
\input psfig

\title{\LARGE \bf
Theoretical Modelling of Langmuir Monolayers
}
\author{\bf
F. Schmid, C. Stadler, H. Lange 
\\
{\em
Johannes Gutenberg Universit\"at Mainz,
D55099 Mainz, Germany
}}
\date{\today}
\maketitle

\bigskip
{\bf Abstract} -

We study  coarse grained, continuum models for Langmuir monolayers by
self consistent field theory and by Monte Carlo simulations.
Amphiphilic molecules are represented by stiff chains of monomers
with one end grafted to a planar surface. In particular, we 
discuss the origin of successive fluid-fluid transitions, the possible 
origin of tilt order and the factors which determine the direction of tilt. 

\vfill
\noindent
Paper presented at the {\em 9th International Conference on Surface and
 Colloid Science}, Sofia, Bulgaria, 6-12 July 1997.
\newpage

\section{Introduction}

In this contribution, we present theoretical model calculations for monolayers 
of amphiphiles spread at the air/water interface, {\em i.e.}, Langmuir
monolayers. Such monolayers have two particularly remarkable properties: 
First, at intermediate temperatures there exist two distinct regions of 
fluid-fluid coexistence -- a ``gas-liquid'' coexistence region and an 
additional region at higher surface coverage, where ``liquid condensed'' 
domains are present in a ``liquid expanded'' (LE) environment. The transition 
from the liquid expanded state to the liquid condensed state is not first 
order in a strict sense. Long range electrostatic interactions between 
the amphiphile head groups prevent macroscopic phase separation, and lead to 
the formation of a superstructure of ordered domains instead. However, these 
domains are of mesoscopic size ($\mu m$), and the transition can be considered
first order on smaller length scales\cite{rev1}. The second noteable feature 
of Langmuir monolayers is a complex polymorphism of phases on the condensed 
side of the phase diagram, which differ from each other in tilt order, 
positional order and orientational order of the backbones of the 
chains\cite{rev2}. Three different types of phases, all of them liquid, can 
coexist with the liquid expanded phase: an untilted phase, and tilted phases 
with the chains tilted in the direction of nearest neighbors (NN), or next 
nearest neighbors (NNN). 

We shall focus on two questions here: The origin of the 
first order transition to the liquid expanded phase, and the factors
which determine the occurrence and the direction of collective
tilt. To this end, we study simplified models of endgrafted
stiff chains using mean field methods and Monte Carlo simulations.
In the next two sections, we discuss mean field arguments with the
aim to gain some insight into these problems. Then, we present
some results of Monte Carlo simulations.

\section{Transition Liquid Expanded/Liquid Condensed}

The amphiphiles are modeled as chains of seven rodlike segments 
(length $l_0$, diameter $A_0$) and one head segment, which is confined to 
a planar surface by a harmonic potential. Our chains are thus relatively
short. Assuming that one segment corresponds to roughly two CH${}_2$ units,
they model in a very idealized way molecules of the size of, {\em e.g.}, 
tetradecanoic acid. 
The chains are made stiff with a bond angle potential 
$u \: k_B T \widehat{U}(1-\cos \theta)$ for the angle 
$\theta$ between adjacent segments. We choose $\widehat{U}(x)$ such that it
has a global minimum at $x=0$ (straight chains) and a local minimum
at $x=1/2$ (chain defects): $\widehat{U}(x)= 25x+34x^2-400x^3+480x^4$.
We note however that qualitatively similar results are obtained with the 
simpler form $\widehat{U}(x)=x$. The adjustable parameter $u$ determines the 
stiffness of the chains. Chain segments interact {\em via} 
repulsive hard core interactions and attractive long range interactions.
These are treated with a density functional formalism, discussed in
detail in ref. \cite{me3}. In a mean field treatment, we also need to
account for the anisotropy of the chains explicitly. We do so on the
level of the second virial coefficient by adding an orientation dependent 
term: $B_2^{anis.}(\Phi) = v \: \frac{5}{16 \pi} (3 \cos^2 \Phi-1)$,
where $\Phi$ is the angle between two interacting segments. The
parameter $v$ thus describes the anisotropy per segment of a chain.

The problem is solved using self consistent field theory. In short,
segment density distribution functions are calculated for single
chains in an inhomogeneous external field, which is in turn
determined self consistently from the density distribution functions.
One obtains density profiles (Figure 1) and free energies.

Figure 2 shows a resulting phase diagram in the plane of chain
anisotropy versus molecular area. One finds coexistence of a
liquid phase and a gas phase (not shown in the figure), 
and a region of phase coexistence between two untilted liquid 
phases, bounded by a critical point at low chain anisotropy and a triple 
point at high chain anisotropy. Almost the same diagram is obtained 
when varying the chain stiffness at fixed chain anisotropy 
Since both the effect of segment interactions and the chain stiffness
go down with increasing temperature, the $v$ or $u$ axis can also
be interpreted as temperature axis. 

Hence the phase behavior turns out bo be mainly driven by the
chain stiffness and the chain anisotropy. One can infer the origin 
of the phase transition: It is the result of a competition between the 
conformational entropy of the chains, which favors a disordered expanded
state, and the tendency of the chains to pack parallel to each other,
which favors the more compact condensed state. The expanded phase
is thus dominated by chain ``melting'', and the condensed state by
collective chain alignment \cite{me3,me1}.
We note that our mean field treatment does not include the possibility of 
hexatic order. Therefore the coexistence region ends in a critical point.
If the condensed phase has hexatic order, the coexistence region
ends in a multicritical point, and one gets a line of continuous 
Kosterlitz-Thouless type transitions at higher temperatures.

\section{Collective Tilt}

Collective tilt can be induced for a number of reasons \cite{me3}.
In most cases, the dominant mechanism is presumably a simple 
mismatch between the sizes of the head segment and the tail segments. 
When the heads are large, the best way for the tails to maintain
optimal packing is to tilt collectively in one direction. This effect can 
already be modelled by a simple system of rigid rods attached to a head group,
which is confined to a planar surface (Figure 3). The interaction
energy of two rods of length $L$ at grafting distance $\vec{r}$, with 
tilt direction $\vec{e}$, is given by
\begin{equation}
E(\vec{r},\vec{e}) = 
V_h(r) + \frac{1}{\sigma^2}\int_0^L \int_0^L dl \: dl' \: V_{LJ}(d),
\end{equation}
where $d = |\vec{r} + (l-l') \vec{e}|$ is the distance between two
infinitesimal elements on the rod. $V_{LJ}$ is a truncated
Lennard Jones potential 
\begin{equation}
\label{vlj}
V_{LJ}(d) = 
\epsilon ((\frac{\sigma}{d})^{12}-2 (\frac{\sigma}{d})^{6}+0.031) 
\quad (d\le 2); \qquad
V_{LJ}(d) = 0 \quad \mbox{otherwise}
\end{equation}
and the head potential $V_h(d)$ is purely repulsive
\begin{equation}
\label{vh}
V_{h}(r) = \epsilon ((\frac{\sigma_h}{r})^{12}-2 (\frac{\sigma_h}{r})^{6}+1) 
\quad (d\le \sigma_h); \qquad
V_{h}(r) = 0 \quad \mbox{otherwise}
\end{equation}
with the head size $\sigma_h$. Figure 4 shows the state of lowest
energy (``ground state'') for such a system. As a function of head
size or surface pressure, one finds a sequence of tilting
transitions: No tilt (U) at small head size or high surface
pressure, tilt towards next nearest neighbors (NNN) at intermediate
head size or surface pressure, and tilt towards nearest neighbours (NN)
at large head size and low surface pressure.

Hence not only the presence of tilt, but also the direction of tilt
appears to be determined by the size of the head groups. The latter result
might seem somewhat unexpected. It results from an interplay of
surface tension and volume packing effects in the monolayer. An
intuition can be gained from closer inspection of Figure 3. 
As the chains tilt, the head lattice gets distorted in the 
direction of tilt. In case of tilt towards next nearest neighbors
(b), the chain can thus optimize the distance to the four direct
neighbors of type (ii); in case of tilt towards nearest neighbors (a),
only the distance to the two neighbors (i) can be optimized.
Hence the chain packing in the bulk is better for next nearest
neighbor tilt. On the other hand, due to the tilt induced lattice distortion
all direct neighbors are pushed apart in the case of tilt towards nearest
neighbors, and only the neighbors of type (ii) in the case of tilt
towards next nearest neighbors. Hence large heads are more comfortable
in an environment of chains tilted towards nearest neighbors. 

The argument can be translated into a simple approximate expression for
the free energy \cite{me2}. According to this simplified (analytical)
treatment, one expects a transition from the untilted state to a state with
tilt towards next nearest neighbors as soon as the head size $\sigma_h$ 
becomes larger than the effective chain diameter $r_t$
($r_t \approx 0.93 \sigma $ in our case). The direction of tilt is expected
to switch towards nearest neighbors at 
\begin{equation}
\label{rr}
\sigma_h - r_t = 3.93 r_t L/(2 \kappa (\Sigma+\Pi)
\end{equation}
where $L$ is the chain length, $\kappa$ is the volume compressibility of 
the chains, $\Sigma$ is the surface tension and $\Pi$ is the surface pressure
(cf. Figure 4).

\section{Monte Carlo simulations}

We have performed Monte Carlo simulations of systems of stiff bead-spring 
chains, with one head bead confined to a planar surface \cite{frank}. 
Tail beads which are not direct neighbors in the chain interact {\em via} the 
Lennard-Jones potential (\ref{vlj}), and head beads interact with 
the repulsive potential (\ref{vh}). The beads are connected by bonds of
length $b$, subject to the pseudoharmonic potential
$V_{bl}(b)=-2 \ln(1-25 (b-b_0)^2)$, where the optimum bond length is chosen
$b_0=0.7 \sigma$. The chains are made stiff
with a bond angle potential $U(\theta)=10 (1-\cos \theta)$. In our
simulations, chains contain one head bead and six tail beads. System sizes
are typically 144 chains. Figure 5 shows a configuration snapshot in the 
disordered phase.

A full account of our simulations with detailed discussions of the results will
be published elsewhere (\cite{coming}, see also \cite{me4}). Here, we just
show the resulting phase diagram for heads of size $\sigma_h=1.2 \sigma$
(Figure 6). We find a number different condensed phases, tilted and untilted, 
and a transition to a disordered phase which is first order for temperatures
below $T \approx 3 \epsilon/k_B$. At small surface pressure, the
tilted phase is tilted towards next nearest neighbors. As the pressure
is increased, an additional phase with tilt towards nearest neighbors
emerges. Hence our simulations basically confirm the physical picture,
which we have obtained by mean field arguments.

\section{Discussion}

To summarize, we have shown that the theoretical study of simplified
models for monolayers of amphiphiles helps to clarify some of the physics 
of phase transitions in these systems. In particular, we have
demonstrated that the first order transition between the liquid expanded
phase and the liquid condensed phase is essentially driven by the
chains. The condensed phase is characterized by chain alignment, {\em i.e.},
efficient chain packing, whereas the expanded phase is dominated by chain 
disorder. Furthermore, we have studied the phenomenon of collective tilt and
established a relationship between the size of the head groups, the
surface pressure, and the direction of the tilt. 

Our results are consistent with a number of experimental observations. 
For example, it has been shown that the liquid expanded phase disappears 
when the carbon chains are perfluorinated, {\em i.e.}, made much stiffer.
It can be brought back into existence by the introduction of flexible 
hydrocarbon spacers\cite{barton}. Thus chain flexibility turns out to be
important in the liquid expanded phase. 

Our claim, that the phase behavior is mainly driven by the chains, is 
furthermore supported by the ``temperature effect''\cite{bibo}, according to 
which the addition of two CH${}_2$ units to the chain has an effect similar to 
reducing the temperature by 10-20${}^0$C. 
We have not calculated explicitly the influence of the chain length on the 
phase behavior. However, one can argue, within our model, that increasing the 
chain length roughly corresponds to making the chains more flexible and
lowering the temperature simultaneously (see Ref. \cite{me3} for details).  

Finally, we have provided a simple explanation for the sequence of
tilt transitions as a function of surface pressure in fatty acid 
monolayers \cite{rev2}. The relation between head size and tilt direction 
also accounts for the fact that tilt towards
nearest neighbors can be suppressed by increasing the pH of the 
subphase \cite{shih}, or by replacing the COOH head groups in part by
smaller alcohol head groups \cite{fischer,teer}

\section{Acknowledgments}

We have benefited from discussions with M. Schick, K. Binder, F.M. Haas
and K. Binder. C.S. is supported by the Graduiertenkolleg on
supramolecular systems in Mainz.

\newpage
\pagestyle{empty}

\section*{Figure Captions}
\parbox{14cm}{
\lettersize

\begin{description}

\item[Figure 1:] Density profiles obtained from self consistent field 
calculations in the direction $z$ perpendicular to the interface for 
different molecular areas $A$. Long and short dashed lines show the center of 
mass densities of tail and head segments, respectively. The solid line shows a 
coarse grained density profile, which accounts for the finite extension of 
the segments. Units are: $z$, the segment length $l_0$; $A$, the segment
diameter $A_0$; and densities, $1/(l_0 A_0)$.
Parameters are $u=2$ and $v=13.7$.

\item[Figure 2:] Phase diagram obtained from self consistent field
calculation in the plane of anisotropy per segment $v$ and
molecular area $A$ (in units of the segment diameter $A_0$)
at chain stiffness $u=2$ (after Ref. \cite{me3}).

\item[Figure 3:] Side view of the rigid rod model (top) and top view
of the head lattice (bottom). (i) and (ii) mark different types
of direct neighbors.

\item[Figure 4:] Zero temperature phase diagram of the rigid rod
model in the plane of head size $\sigma_h$ (in units of $\sigma$) vs. surface 
pressure $\Pi$ (in units of $\epsilon/k_B \sigma^2$) at rod length 
$L=5 \sigma$. Dashed lines indicate the prediction of eqn.  (\ref{rr}) with 
$\Sigma = 6.7 \epsilon/k_B \sigma^2$ (after Ref. \cite{me2}).

\item[Figure 5:] Configuration snapshot in the disordered phase.
Parameters are: Pressure $\Pi = 40\epsilon/k_B \sigma^2$, temperature
$T=4 \epsilon/k_B$ and head size $\sigma_h=1.1 \sigma$.
(from Ref. \cite{me4}).

\item[Figure 6:] Phase diagram obtained from Monte Carlo simulations in the 
plane of pressure $\Pi$ (in units of $\epsilon/k_B \sigma^2$) vs. temperature 
$T$ (in units of $\epsilon/k_B$) for head size $\sigma_h = 1.2 \sigma $.
The transition to the disordered phase is first order up to 
$T \approx 3 \epsilon/k_B$
and possibly becomes second order at higher temperatures.

\end{description}
}

\newpage

\figpage{1}{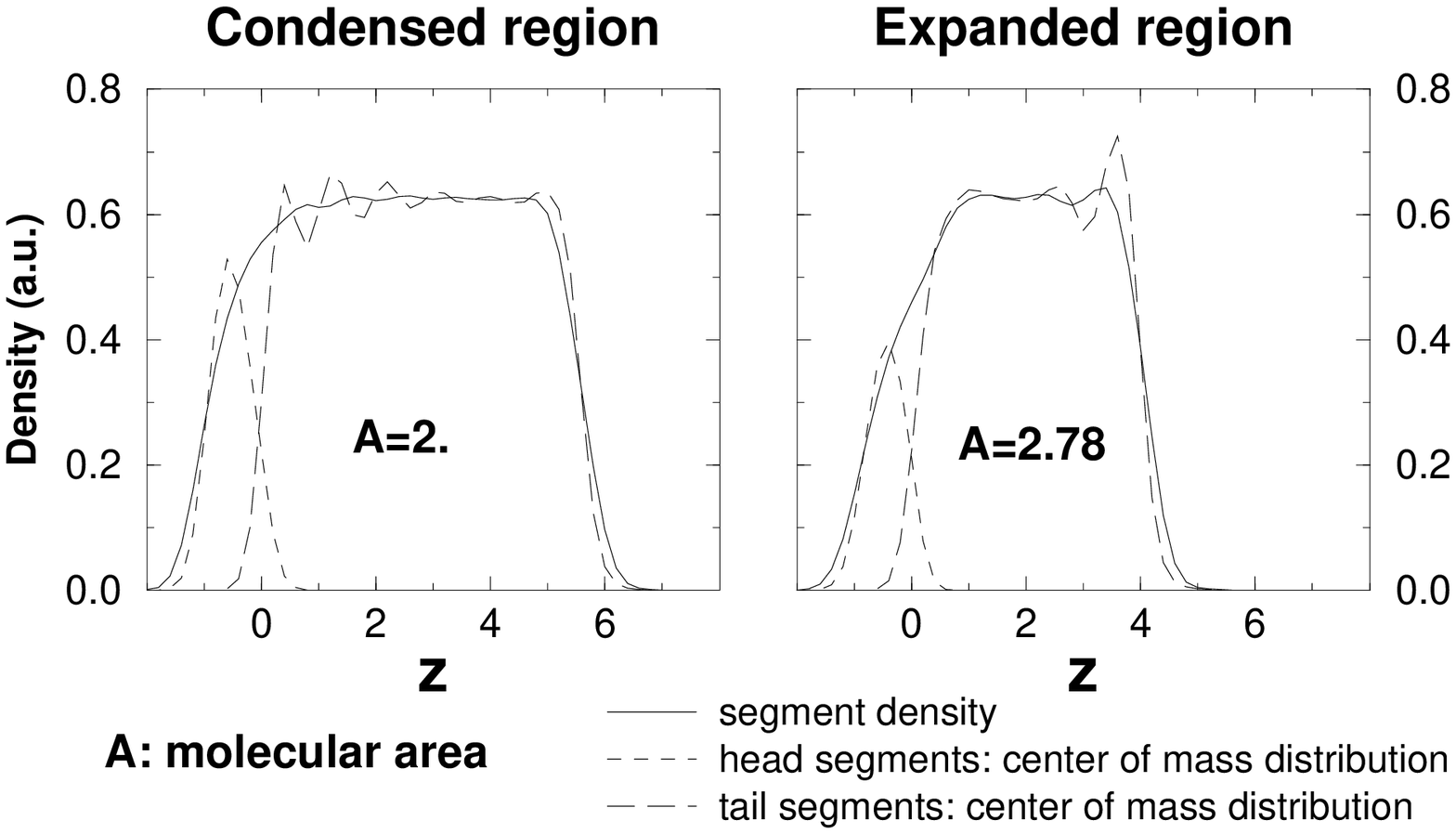}

\figpage{2}{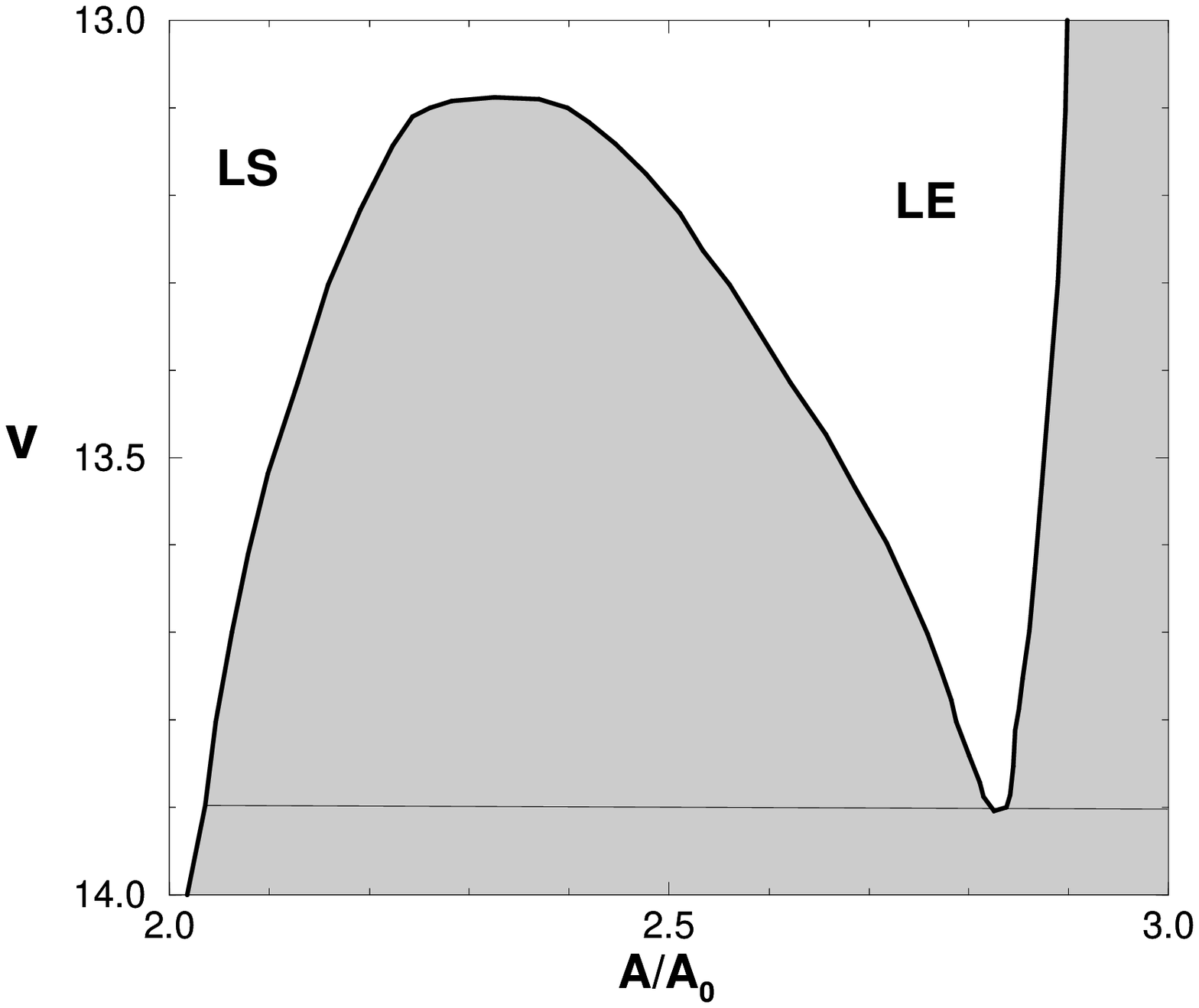}

\xfigpage{3}{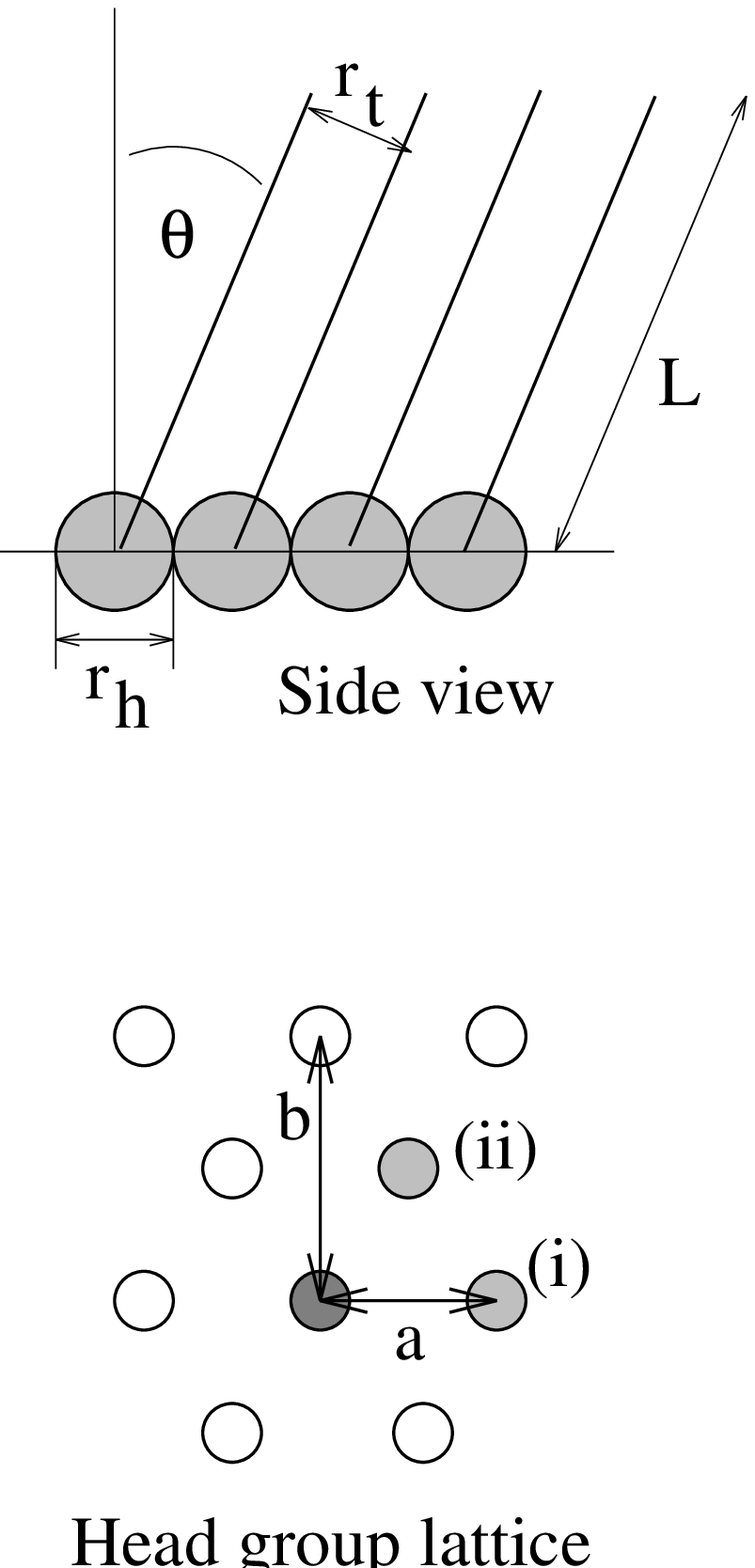}

\figpage{4}{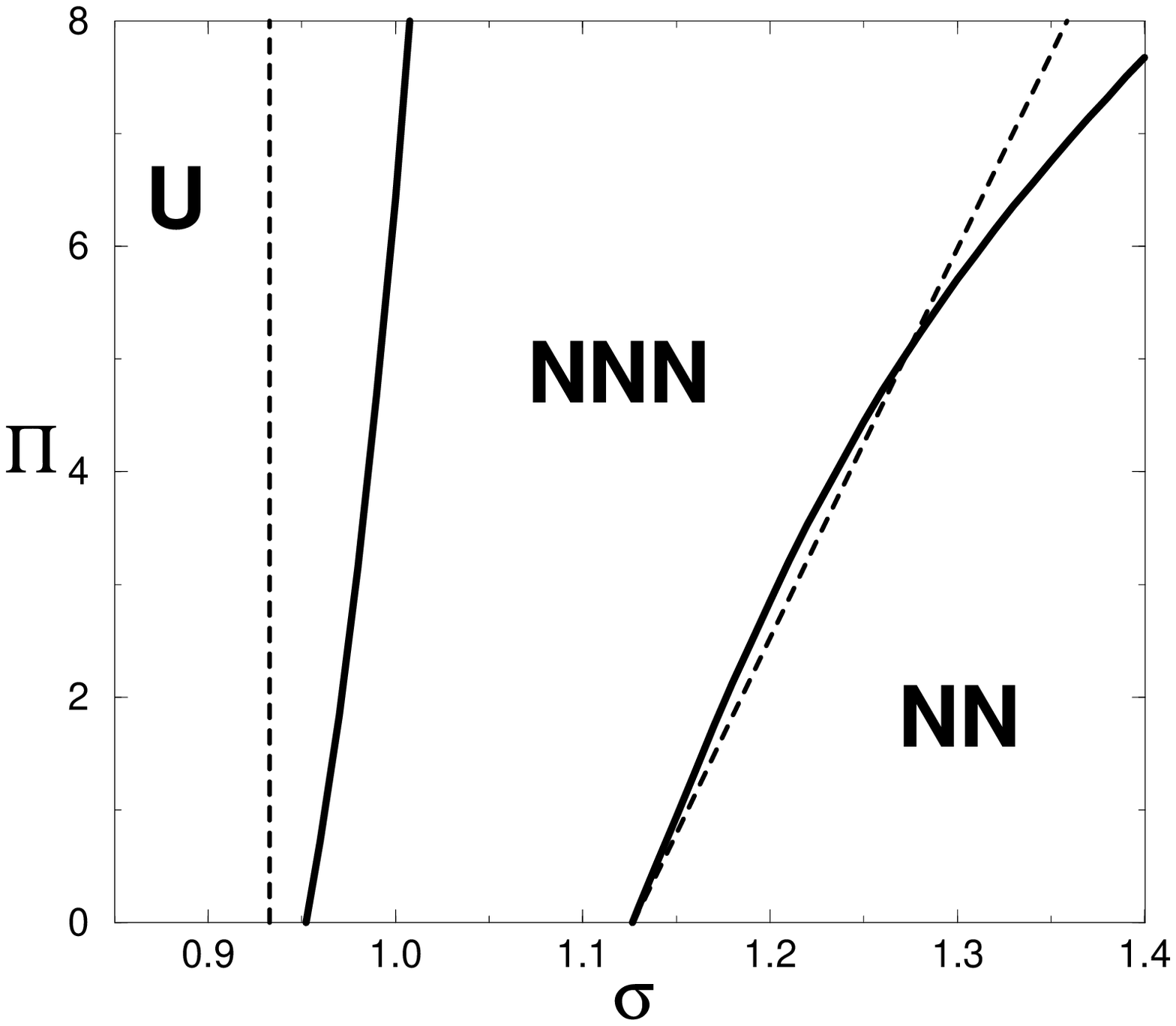}

\confpage{5}{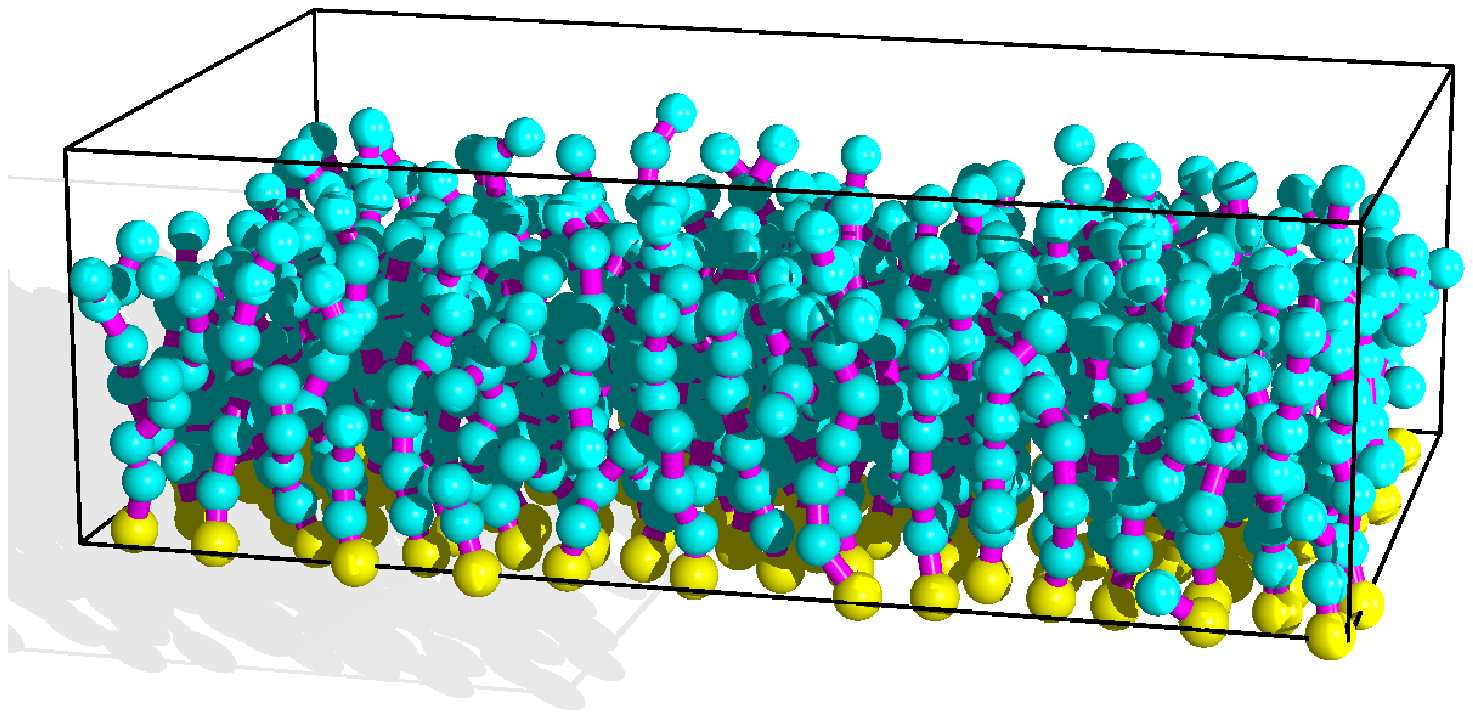}

\figpage{6}{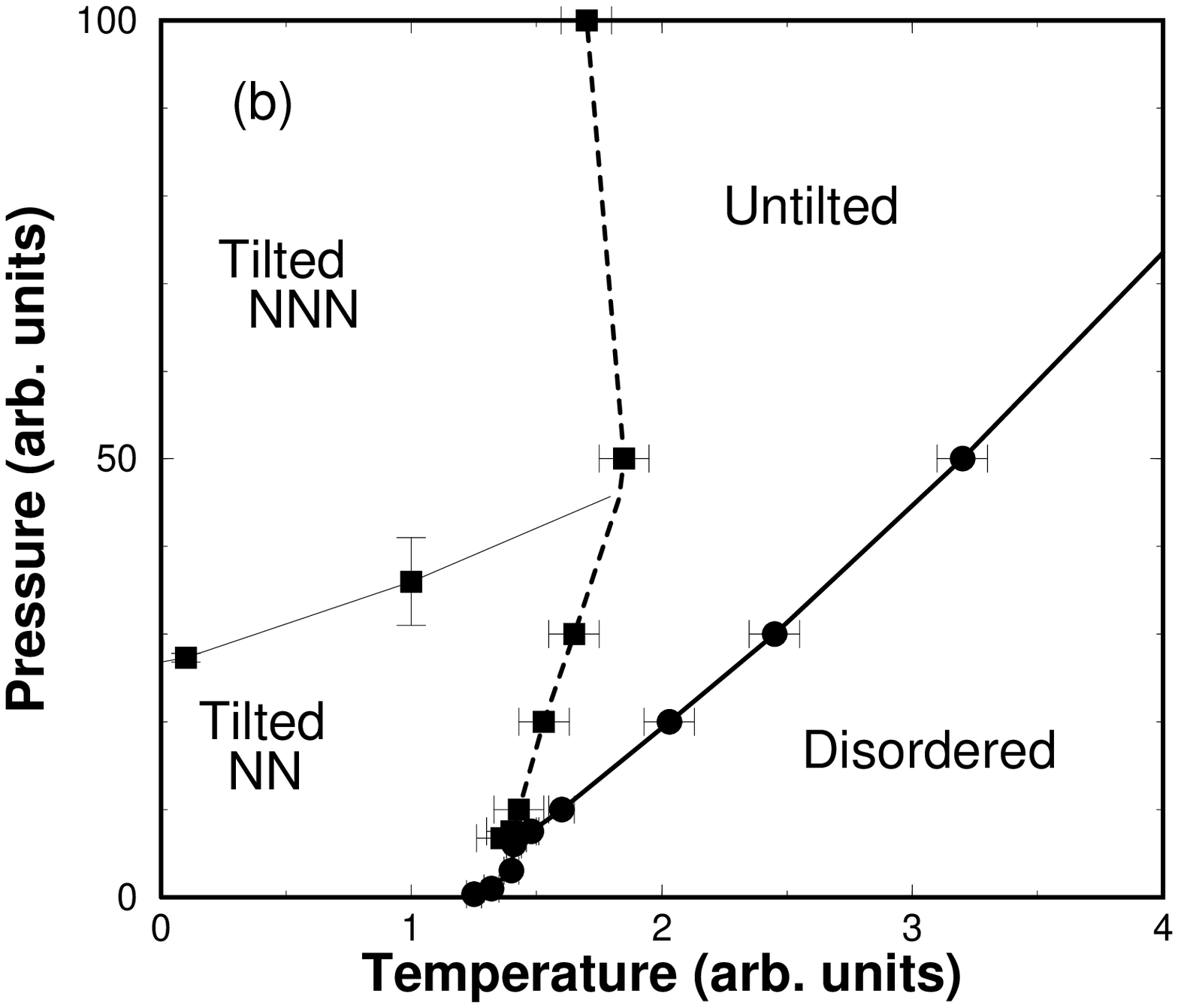}

\end{document}